\documentstyle[aps,12pt]{revtex}
\begin{document}
\begin{center}
\thispagestyle{empty}

{\normalsize\begin{flushright}
CERN-TH-99-309\\ 
HD-THEP-99-47\\[20ex] 
\end{flushright}}

{\large \bf Fluctuations from dissipation 
            in a hot non-Abelian plasma   }\\[6ex]

{Daniel F. Litim 
\footnote{E-Mail: D.Litim@thphys.uni-heidelberg.de}${}^{,a}$ 
and Cristina Manuel 
\footnote{E-Mail:  Cristina.Manuel@cern.ch}${}^{,b}$}
\\[4ex]
{\it ${}^a$Institut f\"ur Theoretische Physik, Philosophenweg 16, 
D-69120 Heidelberg, Germany.\\[2ex]
${}^b$Theory Division, CERN, CH-1211 Geneva 23, Switzerland.}
\\[10ex]
 
{\small \bf Abstract}\\[2ex]
\begin{minipage}{14cm}{\small
We consider a transport equation of the Boltzmann-Langevin type for
non-Abelian plasmas close to equilibrium to derive the spectral functions
of the underlying microscopic fluctuations from the entropy. The
correlator of the stochastic source is obtained from the dissipative
processes in the plasma. This approach, based on classical transport
theory, exploits the well-known link between a linearized collision
integral, the entropy and the spectral functions. Applied to the
ultra-soft modes of a hot non-Abelian (classical or quantum) plasma, the
resulting spectral functions agree with earlier findings obtained from the
microscopic theory. As a by-product, it follows that B\"odeker's effective
theory is consistent with the fluctuation-dissipation theorem.}
\end{minipage}
\end{center}

\newpage
\pagestyle{plain}
\setcounter{page}{1}
It has been recognized that dynamical properties of
(non-perturbative) quasi-particle excitations in non-Abelian plasmas
can be described very efficiently by means of effective transport
equations. A prominent recent example is given by B\"odeker's effective
classical theory for the ultra-soft modes in a hot non-Abelian plasma
close to equilibrium \cite{Bodeker}, which corresponds to a transport
equation of the Boltzmann-Langevin type. In \cite{LM}, a general 
procedure has been presented, based on classical coloured point particles,
to obtain effective transport equations from the microscopic theory after 
integrating-out the fluctuations about the mean values, and taking the
Gibbs ensemble average in phase space. On the one-loop level, the same mean
field equations of \cite{LM} have been obtained recently within a 
many-particle world line formalism \cite{JJVW} (see also
\cite{Pisarski}). The collision integral and the source of 
stochastic noise of \cite{Bodeker} have been obtained from \cite{LM} 
to leading order in a weak coupling expansion, and at logarithmic accuracy. 
It was also realized that the dynamical equations 
are the same for the classical and the quantum plasma, changing only in 
the value of the Debye mass \cite{LM}. 
Other approaches to obtain the collision term of \cite{Bodeker}
have been reported as well \cite{Arnold,BI,Valle}.

In the present article we consider, based on classical transport theory, a
generic Boltzmann-Langevin equation for the
one-particle distribution function ${f}(x,p,Q)$, given as 
\begin{equation}    \label{EffTrans}
p^\mu\left(\frac{\partial}{\partial x^\mu}
- g f^{abc}A^{b}_\mu Q^c\frac{\partial}{\partial Q^a}
-gQ_aF^{a}_{\mu\nu}\frac{\partial}{\partial p_\nu}\right) {f}(x,p,Q)
= C[{f}](x,p,Q) +\zeta(x,p,Q) \ . 
\end{equation}
Here, the variables $Q$ describe the non-Abelian colour charges.
The transport equation contains an effective collision term $C[f]$ and
an associated source for stochastic noise $\zeta$. The SU($N$) gauge fields 
appearing
in the above equation are self-consistent, that is, generated by the
same particles of the plasma. The Yang-Mills equation are
\begin{equation} \label{YM}
\left(D_\mu F^{\mu \nu} \right)_a = J^\nu_a (x)  
\equiv g \sum_{\hbox{\tiny helicities}\atop\hbox{\tiny species}} 
    \int dP \,dQ\, Q_a\, p^\nu\,f(x,p,Q) \ ,
\end{equation}
where the momemtum measure reads
$dP = d^4p 2\Theta(p_0) \delta(p^2-m^2)$, 
and the colour measure $dQ$ was defined in \cite{LM}.
The current in (\ref{YM}) is covariantly conserved.
We work in natural units $c=\hbar=k_B=1$, unless 
otherwise specified. From now on we will omit
the sum over different species of particles and helicities.

In the collisionless limit
$C=\zeta=0$, the above set of transport equation reduces to those
introduced by Heinz \cite{Heinz}. In the general case however,
the r.h.s.~of (\ref{EffTrans}) does not vanish due to effective
interactions (collisions) in the plasma, resulting in the term
$C[f]$. In writing (\ref{EffTrans}), we have already made the assumption that
the one-particle distribution function $f$ is a fluctuating quantity. 
This is quite natural having in mind that $f$ describes a 'coarse-grained' 
microscopic distribution function for coloured point particles, and
justifies the presence of the stochastic source $\zeta$ in the transport 
equation. For non-charged particles, a similar kinetic equation 
has already been considered in \cite{Bixon} (see also \cite{CH}, where 
stochastic noise is introduced to a Schwinger-Dyson approach). 

Given the stochastic dynamical equation (\ref{EffTrans}), the question
raises as to what can be said on general grounds about the spectral
functions of $f$ and $\zeta$. Here, we shall assume that the
dissipative  processes are known close to equilibrium, but no further
information is given regarding the  underlying fluctuations. This way
of proceeding is complementary to \cite{LM}, where  the r.h.s.~of
(\ref{EffTrans}) has been obtained from correlators of the microscopic
statistical fluctuations.  We then show that the spectral function of
the fluctuations and the noise correlator close to equilibrium can be
obtained from the knowledge of the entropy of the plasma, and from the
dissipative term in the effective transport equation. This gives a
well-defined prescription as to how the correct source for noise can
be identified without the detailed knowledge of the underlying
microscopic dynamics responsible for the dissipation. The basic idea
behind this approach relies on the essence of the
fluctuation-dissipation theorem (FDT). While this theorem is more general,
here we will only discuss the close to equilibrium situations.
According to the FDT if a fluctuating system remains close to
equilibrium, then the dissipative process  occurring in it  are
known. Vice versa, if one knows the dissipative process in the system,
one can describe the fluctuations without an explicit knowledge of the
microscopic structure or processes in the system. The cornerstone of
our  approach is the entropy of the fluctuating system, which serves
to identify the thermodynamical forces, and leads to the spectral
function for the deviations from the non-interacting equilibrium.

Before entering into the discussion of plasmas, we will
illustrate this way of proceeding by  reviewing the simplest
setting of  classical linear dissipative systems \cite{Landau}.  A
generalization to the more complex case of non-Abelian plasmas will
then become a  natural step to perform.   We consider a classical
homogeneous system described by a set of variables $x_i$, where $i$ is
a discrete index running from $1$ to $n$.  These variables are
normalized in such a way that their mean values at equilibrium
vanish. The entropy of the system is a function of the quantities
$x_i$, $S(x_i)$. If the system is at equilibrium, the entropy reaches
its maximum, and thus $(\partial S/ \partial x_i)_{\rm eq} = 0$,
$\forall i$. If the system is taken slightly away from equilibrium,
then one can expand the difference $\Delta S =S- S_{\rm eq}$, where
$S_{\rm eq}$ is the entropy at equilibrium, in powers of $x_i$. If we
expand up to quadratic order, then
\begin{equation} 
\Delta S = \frac 12 \left(
\frac{ \partial^2 S}{ \partial x_i \partial x_j} \right)_{\rm eq} x^i x^j 
\equiv  - \frac 12 \beta_{ij} x^i x^j  \ .  
\end{equation} 
The matrix $\beta_{ij}$ is symmetric and
positive-definite, since the entropy reaches a maximum at equilibrium.
The thermodynamic forces $F_i$ are defined as the gradients of $\Delta S$
\begin{equation}
F_i = - \frac{ \partial \Delta S}{ \partial x_i}  \ .
\end{equation} 
For a system close to equilibrium  the thermodynamic forces are linear
functions of $x_i$, $F_i =  \beta_{ij} x^j$.
If the system is at equilibrium, the thermodynamic forces vanish. 
In more general situations the variables 
$x_i$ will evolve in time.
The time evolution of these variables is given as functions
of the thermodynamical forces. In a close to equilibrium case
one can expect that the evolution is linear in the forces
\begin{equation}
\frac{d x^i}{dt} = - \gamma^{ij} F_j + \zeta^i \ ,
\end{equation} 
which, in turn, can be expressed as
\begin{equation}
\frac{d x^i}{dt} = - \lambda^{ij} x_j + \zeta^i \ ,
\end{equation} 
The first term in the r.h.s. of the above equation describes the 
mean regression of the system towards equilibrium,  while the second term
is the source for stochastic noise.
The quantities $\gamma^{ij}$ are known as the  kinetic coefficients, 
and it is not difficult to check that
$\gamma_{ij}= \lambda_{ik} \beta^{-1}_{kj}$. 
From the value of the coefficients $\beta_{ij}$ one can deduce the 
equal time correlator
\begin{equation}
\left\langle x_i (t) x_j (t)\right\rangle =  \beta^{-1}_{ij} \ ,
\label{invbet}
\end{equation}
which is used to obtain Einstein's law
\begin{equation}
\left\langle x^i (t) F_j (t)\right\rangle = \delta^i _j  \ .
\label{einstein}
\end{equation}
After taking the time derivative of (\ref{einstein}),  
assuming that the noise is white and Gaussian
\begin{equation}
\left\langle \zeta^i (t) \zeta^j (t') \right\rangle = \nu^{ij} \delta(t-t') \ ,
\end{equation}
we find that the strength of the noise self-correlator $\nu$ is determined
by the dissipative process
\begin{equation}
\nu^{ij} = \gamma^{ij} + \gamma^{ji} \ ,
\label{simpFDT}
\end{equation}
which is the FDT relation we have been aiming at. 

We now come back to the case of a non-Abelian plasma and
generalize  the above discussion to the case of our concern.  We will
consider the non-Abelian plasma as a linear dissipative system, assuming
that we know the collision term in the transport equation. In order to
adopt the previous reasoning, we have to identify the dissipative
term in the transport equation, and to express it as a function of the
thermodynamical force obtained from the entropy. The deviation from the
equilibrium distribution is given here by
\begin{equation}
\Delta f(x,p,Q) = {f}(x,p,Q) - f_{\rm eq}(p_0)\ ,
\end{equation} 
and replaces the variables $x_i$ discussed above. 
%The deviation $\Delta f$ goes as ${\cal O}(g)$ to leading order 
%in a small gauge coupling expansion. 
The entropy flux density for classical plasmas is given as 
\begin{equation} \label{Scl}
S_\mu (x) = - \int dP dQ\, p_\mu \, f(x,p,Q) 
\left( \ln{( f(x,p,Q) h^3)} -1 \right) \ ,
\end{equation}
where $h$ is an arbitrary constant such that $f(x,p,Q) h^3$ is dimensionless.
The $\mu =0$ component of (\ref{Scl}) gives the entropy density 
of the system.
The entropy itself is then obtained as $S = \int d^3x \, S_0 (x)$.

We shall now assume that the deviation of the mean particle number
from the equilibrium one is small within a coarse-graining volume. 
This can always be arranged for at small gauge coupling 
$g\ll 1$.\footnote{More precisely, in the close-to-equilibrium plasma we 
consider $f$ as being coarse-grained over a Debye volume. This entails that 
({\it i}) the particle number fluctuations within a Debye volume 
   are parametrically 
   suppressed by powers of the plasma parameter (that is, ultimately, by $g$), 
({\it ii}) the two- and higher-particle distribution functions are suppressed 
     as opposed to $f$, and 
({\it iii}) typical relaxation scales or times are not affected by the 
      coarse-graining.} 
We then obtain $\Delta S$ just by expanding the expression of the 
entropy density in powers of $\Delta f$ up to quadratic order. It is 
important to take into account that we will consider situations where 
the small deviations from equilibrium are such that both the particle 
number and the energy flux remain constant, thus
\begin{eqnarray}
\int  dP dQ \, \Phi(p)\,      \Delta f(x,p,Q) & = & 0 \ , 
\quad {\rm for}\quad \Phi(p)=p_0,\, p_0p_\mu \, .
\end{eqnarray}
Under those assumptions, one reaches to
\begin{eqnarray}
\Delta S_0 (x) & = &   
- \int dP dQ \, p_0 \,
\frac{(\Delta f(x,p,Q))^2}{2 f_{\rm eq}(p_0)} \nonumber \\
               & = &  
- \int d^3 p \, dQ   
\frac{(\Delta f(x,{\bf p},Q))^2}{2 f_{\rm eq}({\omega_p} )} \ ,
\label{S0}
\end{eqnarray}
where in the last equality we have taken into account the mass-shell 
condition, with $p_0 = \omega_p =\sqrt{{\bf p}^2 + m^2}$. 
Without loss of generality, we will consider from now on the case of massless particles,
so $\omega_p = p = |{\bf p}|$.

The thermodynamic force 
associated to $\Delta f$ is defined from the entropy as
\begin{equation}    \label{F}
F(x,{\bf p},Q)  
= - \frac {\delta \Delta S}{\delta \Delta f(x,{\bf p},Q)} 
=  \frac{\Delta f(x,{\bf p},Q)}{f_{\rm eq}(p)}  \ .
\end{equation}
We now linearize the transport equation (\ref{EffTrans}) and express 
the collision integral close to equilibrium in terms of the 
thermodynamical force. Dividing (\ref{EffTrans}) by $p_0$ and imposing 
the mass-shell constraint, we find
\begin{equation}  \label{linearizedBeq}
   v^\mu D_\mu \Delta f 
-g v^\mu Q_aF^{a}_{\mu 0}\frac{d f_{\rm eq}}{d p} 
=   C[\Delta f](x,{\bf p},Q) 
  + \zeta(x,{\bf p},Q) \ , 
\end{equation}
where $v^\mu = p^\mu/p_0 =(1, {\bf v})$, with ${\bf v}^2 =1$.
We also introduced the shorthand $D_\mu \Delta f\equiv 
(\partial_\mu -g f^{abc}A_{\mu,b}Q_c\partial^Q_a)\Delta f$ \cite{LM}.
It is understood that the collision integral has been linearized,
and we write it as
\begin{equation}\label{Cf} 
C[\Delta f] (t,{\bf x},{\bf p},Q) =  
\int  d^3 x' d^3 p' \, dQ'\,
{K}({\bf x},{\bf p},Q;{\bf x}',{\bf p'},Q') 
\Delta f(t,{\bf x}',{\bf p}',Q')\, ,  
\end{equation} 
with $t\equiv x_0$. For simplicity, we take the collision integral 
local in time, but unrestricted elsewise.\footnote{Of course, 
gauge invariance imposes further conditions on both the collision 
term and the noise. However, these constraints are of no relevance 
for the present discussion.}  
According to the FDT, the source of stochastic noise has to obey 
\begin{equation} 
\left\langle  \zeta(x,{\bf p},Q)  
          \zeta(x',{\bf p}',Q') \right\rangle  
=  
- \left(
 \frac{\delta C[F](x,{\bf p},Q)}{\delta F(x',{\bf p}',Q')}  
+\frac{\delta C[F](x',{\bf p}',Q')}{\delta F(x,{\bf p},Q)}\ 
\right)
\end{equation} 
in full analogy to (\ref{simpFDT}).  With the knowledge of the
thermodynamical force (\ref{F}) and the linearized collision term
(\ref{Cf}) we arrive at 
\begin{equation}\label{noisecorrelator} 
\left\langle \zeta(x,{\bf p},Q)  
         \zeta(x',{\bf p}',Q') \right\rangle  
=  
- \left( 
 {f_{\rm eq}({p})}{K}({\bf x},{\bf p},Q;{\bf x}',{\bf p'},Q') 
              + {\rm sym.} \right) \delta ( t-  t')\ .  
\end{equation}
Here, symmetrisation in 
$({\bf x},{\bf p},Q)  \leftrightarrow ({\bf x}',{\bf p}',Q')$
is understood.

Notice that we can derive the equal time correlator for the deviations
from the equilibrium distribution simply from the knowledge of the entropy 
and the thermodynamical force, exploiting Einstein's law in full analogy 
to the corresponding relation (\ref{invbet}). 
Using (\ref{F})  we find
\begin{equation} \label{outofeqcorr} 
\left\langle \Delta f(x,{\bf p},Q)  
         \Delta f(x',{\bf p}',Q') \right\rangle_{t=t'}  
=
f_{\rm eq} (p) \delta^{(3)} ({\bf x} - {\bf x}')  
               \delta^{(3)} ({\bf p} - {\bf p}')  
               \delta( Q - Q')  \ .  
\end{equation} 
If the fluctuations
$\Delta f$ have vanishing mean value, then  (\ref{outofeqcorr})
reproduces the well-known result that the correlator  of fluctuations
at equilibrium is given by the equilibrium distribution
function. In order to make contact with the results of \cite{LM}, we
go a step further and consider the case where $\Delta f$
has a non-vanishing mean value to leading order in the gauge
coupling. Splitting  $\Delta f= g {\bar f}^{(1)} + \delta f$ into a
deviation of the mean part  $\langle \Delta f\rangle= g {\bar
f}^{(1)}$ and a fluctuating part  $\langle \delta f\rangle=0$ and
using (\ref{outofeqcorr}), we obtain the equal time
correlator for the fluctuations
$\delta f$ as  
\begin{eqnarray} 
\left\langle \delta f(x,{\bf p},Q)  
         \delta f(x',{\bf p}',Q') \right\rangle_{t=t'}  
&=& 
f_{\rm eq} (p) \delta^{(3)} ({\bf x} - {\bf x}')  
               \delta^{(3)} ({\bf p} - {\bf p}')  
               \delta( Q - Q')
\nonumber \\ && \label{fluccorr} 
- \left. g^2 {\bar f}^{(1)}(x,{\bf p},Q)
             {\bar f}^{(1)}(x',{\bf p}',Q')\right|_{t=t'}  \ .  
\end{eqnarray} 
This result agrees with the correlator obtained in \cite{LM} from the 
Gibbs ensemble average as defined in phase space in the limit where 
two-particle correlations are
small and given by products of one-particle correlators.

Up to now we have dealt with purely classical plasmas. On the same
footing, we can consider the soft and ultra-soft modes in a hot quantum 
plasma. These can be treated classically as their occupation numbers 
are large. The sole effect from their quantum nature reduces to the different
statistics, Bose-Einstein or Fermi-Dirac as opposed to Maxwell-Boltzmann.
The corresponding quantum FDT reduces to an effective classical 
one \cite{Landau,K}. 

Some few changes are necessary to study hot
quantum plasmas. As in \cite{LM}, we change the normalisation of $f$ 
by a factor of $(2\pi\hbar)^3$ to obtain the standard normalisation for 
the (dimensionless) quantum distribution function. Thus, the momentum 
measure is also modified by the same factor,
$dP= {d^4 p} 2\Theta (p_0) \delta(p^2)/{(2\pi\hbar)^3}$ 
for massless particles, and $\hbar =1$. To check the FDT relation
in this case one needs to start with the correct expression for the
entropy for a quantum plasma. The entropy flux density, as a function
of $f(x,p,Q)$, is  given by
\begin{equation} \label{Sq}
S_\mu (x) = - \int dP dQ\, p_\mu \, 
\Big( f \ln{ f} 
       \mp \left( 1 \pm f\right) \ln{(1 \pm f)}  
\Big) \ ,
\end{equation}
where the upper/lower sign applies for bosons/fermions.
From the above expression of the entropy one can compute 
$\Delta S$, and proceed exactly as
in the classical  case, expanding the entropy up to quadratic order in the
deviations from equilibrium. Thus, we obtain the noise correlator 
\begin{equation}\label{noisecorrelatorQ}
\left\langle \zeta(x,{\bf p},Q) 
         \zeta(x',{\bf p}',Q') \right\rangle 
= -
(2\pi)^3 \left(
{f_{\rm eq}({p})}(1 \pm  f_{\rm eq}(p))  
         {K}({\bf x},{\bf p},Q;{\bf x}',{\bf p}',Q')+{\rm sym.}\right)
         \delta (t-t') \ .
\end{equation}
Again, the spectral functions of the deviations from 
equilibrium are directly deduced from the entropy. As a result, we find
\begin{equation} \label{outofeqcorrQ}
\left\langle \Delta f(x,{\bf p},Q) 
         \Delta f(x',{\bf p}',Q') \right\rangle_{t=t'} 
=  
(2\pi)^3 f_{\rm eq} (p)(1 \pm  f_{\rm eq}(p))  
         \delta^{(3)} ({\bf x} - {\bf x}') 
         \delta^{(3)} ({\bf p} - {\bf p}') 
         \delta( Q - Q')  \ .
\end{equation}
Expanding  $\Delta f= g {\bar f}^{(1)} + \delta f$ as above, we obtain the
equal time correlator for $\delta f$, which agrees with the findings of 
\cite{LM} in the case where two-particle distribution functions can be 
expressed as products of one-particle distributions.
 
With the knowledge of the above spectral functions  for the fluctuations
in a classical or quantum plasma one can derive further spectral 
distributions for different physical quantities. In particular, we can 
find the correlations of the self-consistent  gauge field fluctuations
once the basic correlators as given above are known. This is how those
spectral functions were deduced in \cite{LM}.

As a particular example of the above we consider B\"odeker's
effective kinetic equations which couple to the ultra-soft gauge field
modes with spatial extensions $\gg 1/m_D$, where $m_D$ is 
the Debye mass in a non-Abelian plasma close to equilibrium.
The linearized collision integral has been obtained to leading
logarithmic accuracy by several different approaches 
\cite{Bodeker,LM,Arnold,BI,Valle}. To leading order,
they all employ an IR cut-off of the 
order of $gm_D$ for the elsewise unscreened magnetic sector.

We will first concentrate on the classical plasma, 
for particles carrying two helicities. It is most efficient to write 
the transport equation not in terms of the full
one-particle distribution function, but in terms of the current density
\begin{equation} \label{current}
{\cal J}^\rho_a(x,{\bf v})
=  4\pi\, g\, v^\rho \,\int dp\,dQ\, p^2\, Q_a\, \Delta f(x,{\bf p},Q)\, .
\end{equation}
(Notice that $f_{\rm eq}$ gives no contribution to the current.)
The current of (\ref{YM}) follows after integrating over the angles
of ${\bf v}$,
$J_a^\mu (x) = \int \frac{ d\Omega}{4\pi} {\cal J}_a ^\mu(x,{\bf v})$
\cite{LM}.
Expressed in terms of (\ref{current}), the linearized Boltzmann-Langevin 
equation (\ref{linearizedBeq}) becomes 
\begin{equation}    \label{ultracur}
[v^\mu  D_\mu , {\cal J}^\rho](x,{\bf v})
=  -  m^2_D v^\rho v_\mu  F^{\mu0} (x)
   +  v^\rho C[{\cal J}^0](x,{\bf v})
   + \zeta^{\rho}(x,{\bf v})\, .
\end{equation}
where  $m_D$ is the Debye mass \cite{LM}
\begin{equation}\label{mD}
m_D ^2 = - 8 \pi g^2 C_2 \int dp\, p^2 \, \frac{d f_{\rm eq}}{d p} 
\end{equation}
and the constant $C_2$ depends on the 
representation of the coloured particles
\begin{equation}
\int dQ \,Q_a Q_b =C_2 \delta_{ab}\ .
\end{equation}
The linearized collision integral is related to (\ref{Cf}) by 
\begin{equation}    \label{CK}
C[{\cal J}_a^0](x,{\bf v})= 
4 \pi \, g  \int d^3x'\, {d\Omega_{{\bf v}'}}\, dp\, dp'\, dQ \,dQ'\, 
p^2  p'^2 \,Q_a\, 
{K}({\bf x},{\bf p}, Q;{\bf x}',{\bf p'},Q') \,
\Delta f(t,{\bf x}',{\bf p}',Q') 
\end{equation}
and has been obtained explicitly \cite{Bodeker,LM,Arnold,BI,Valle} as
\begin{equation}    \label{C}
C[{\cal J}_a^0](x,{\bf v})=  
 - \gamma\  \int \frac{d\Omega_{{\bf v}'}}{4\pi}\, 
   {\cal I}({\bf v},{\bf v}'){\cal J}_a^0(x,{\bf v}') 
\end{equation}
where the kernel reads
\begin{equation}
{\cal I}({\bf v},{\bf v}')=\delta^{(2)}({\bf v}-{\bf v}')-\frac{4}{\pi}
\frac{({\bf v}\cdot{\bf v}')^2}{\sqrt{1-({\bf v}\cdot {\bf v}')^2}}
\end{equation}
and $\gamma=g^2 N T \ln{(1/g)}/4 \pi$. Comparing (\ref{CK}) with (\ref{C}) we 
learn that only the part of the kernel $K$ which is  symmetric under 
$({\bf x},{\bf p},Q)  \leftrightarrow ({\bf x}',{\bf p}',Q')$ 
contributes in the present case. This part can be expressed as
\begin{equation}
{K}({\bf x},{\bf p}, Q;{\bf x}',{\bf p'},Q') = 
-\gamma \, \frac{{\cal I}({\bf v},{\bf v}')}{4\pi p^2}  
           \delta(p -p') \delta(Q-Q')\delta^{(3)}({\bf x}-{\bf x}')\, .
\end{equation} 
According to our findings above, the self-correlator of the
stochastic source for  the classical plasma obeys
\begin{eqnarray}
\left\langle \zeta^\mu_a (x,{\bf v})\, 
         \zeta^\nu_b (y,{\bf v}') \right\rangle 
&=&  (4 \pi)^2 \, g^2\int dp\, dp'\, dQ\, dQ'\, 
p^2p'^2\,  Q_aQ'_b\,  v^\mu v'^\nu 
   \left\langle \zeta(x,{\bf p},Q)\, \zeta(y,{\bf p}',Q') \right\rangle 
\nonumber \\ &=&
2\,\gamma\,T\,m^2_D 
   \,v^\mu v'^\nu\, {\cal I}({\bf v},{\bf v}') \,
   \delta_{ab}\, \delta^{(4)} (x-y) \ .
\label{corrnoise}
\end{eqnarray}
The sum over the two helicities of the particles has been taken into account.
In order to obtain (\ref{corrnoise}), we have made use of 
(\ref{noisecorrelator}), (\ref{mD}) to (\ref{C}), and of the relation 
$f_{\rm eq}=-T\, df_{\rm eq}/dp$ 
for the Maxwell-Boltzmann distribution.

The quantum plasma can be treated in exactly the same way. To 
confirm (\ref{corrnoise}), we only need to take into account the change 
of normalization as commented above, and the relation
$f_{\rm eq}(1\pm f_{\rm eq})=-T\, df_{\rm eq}/dp$ for the 
Bose-Einstein and Fermi-Dirac distributions, respectively.

Using the explicit expression for the collision integral and the 
stochastic noise it is possible to confirm the covariant conservation 
of the current, $DJ=0$ \cite{Bodeker,LM}.
  
We thus found that the correlator (\ref{corrnoise}) is in full agreement 
with the result of \cite{Bodeker,LM} for both the classical or the
quantum plasma. While this correlator has been obtained in \cite{Bodeker,LM}
from the corresponding microscopic theory, here, it follows solely from the
FDT.  This way, it is established that the effective Boltzmann-Langevin
equation found in \cite{Bodeker} is indeed fully consistent with the 
fluctuation-dissipation theorem. More generally, the important observation
is that the spectral functions as derived here from the entropy and the FDT
do agree with those obtained in \cite{LM} from a microscopic phase space 
average. This guarantees that the formalism of \cite{LM} is consistent with
the FDT.

In the above discussion we have considered the stochastic noise as
Gaussian and Markovian. This is due to the fact that the small-scale
fluctuations (those within a coarse-graining volume) are to leading
order well seperated from the typical relaxation scales  in the
plasma. Within the microscpoic approach  in \cite{LM},  these
characteristics can be understood ultimately as a consequence of an
expansion in a small plasma parameter (or a small gauge coupling).
More precisely, the noise follows to be Gaussian  when higher order
correlators beyond quadratic ones can be neglected.  The Markovian
character of the noise follows because the ultra-soft modes are well
separated from the soft ones, and suppressed in the collision integral
to leading  order. This way, the collision term and the correlator of
stochastic noise are both local in $x$-space. Going beyond  the
logarithmic approximation, we expect from the explicit computation  in
\cite{LM} that the coupling of the soft and the ultra-soft modes makes
the collision term non-local in coordinate space. This non-trivial
memory kernel should also result in a non-Markovian, but still
Gaussian, source for stochastic noise.

The present line of reasoning can in principle be extended to other
approaches.  Using the phenomenological derivation of (\ref{C}) from
\cite{Arnold}, the same arguments as above justify the presence of a
noise source with  (\ref{corrnoise}) in the corresponding Boltzmann
equation \cite{Arnold,BI,Valle}. It might  also be fruitful to follow
a similar line based on the entropy within a quantum field theoretical
language. An  interesting proposal to include self-consistently the
noise within a Schwinger-Dyson approach has been made recently in
\cite{CH}. Along these lines, it might be feasible to derive the
source for stochastic noise directly from the quantum field theory
\cite{BI}.

While we have concentrated the discussion on plasmas close to
equilibrium,  it is known that a fluctuation-dissipation theorem  can
be formulated as well for (stationary and stable) systems
out-of-equilibrium \cite{K}.  It can also be  extended to take non-linear
effects into account \cite{Sitenko}.  Both the out of equilibrium
situations and non-linear effects  can be treated, in principle, with
the general formalism presented in \cite{LM}.

\vskip .5cm

C.M.~thanks the Institute for Nuclear Theory
at the University of Washington for its hospitality and the
Department of Energy for partial support during the completion
of this work.

\end{document}